\documentclass[aps,twocolumn,superscriptaddress,prb,floatfix,british]{revtex4}

\usepackage{amssymb}
\usepackage{amsmath}
\usepackage{babel}
\usepackage{graphicx}
\usepackage[sort&compress]{natbib}
\usepackage{topcapt}
\usepackage{hyperref}

\newcommand{\ket}[1]{\left\lvert #1 \right\rangle}
\newcommand{\bra}[1]{\left\langle #1 \right\rvert}

\newcommand{\GHZ}{\ket{\mathrm{GHZ}}}
\newcommand{\GHZphi}{\ket{\mathrm{GHZ_{\phi}}}}

\newcommand{\Qo}{Q_{1}}
\newcommand{\Qtw}{Q_{2}}
\newcommand{\Qth}{Q_{3}}
\newcommand{\Qf}{Q_{4}}

\newcommand{\Vtw}{V_{2}}

\newcommand{\VH}{V_{\mathrm{H}}}
\newcommand{\mVH}{\langle \VH \rangle}

\newcommand{\Fbare}{f_{\mathrm{c}}}
\newcommand{\Fo}{f_{1}}
\newcommand{\Ftw}{f_{2}}
\newcommand{\Fth}{f_{3}}

\newcommand{\MS}{\mathcal{M}_{\mathrm{S1,2}}}
\newcommand{\MP}{\mathcal{M}_{\mathrm{P1,2}}}

\newcommand{\MSo}{\mathcal{M}_{\mathrm{S1}}}
\newcommand{\MStw}{\mathcal{M}_{\mathrm{S2}}}
\newcommand{\MPo}{\mathcal{M}_{\mathrm{P1}}}
\newcommand{\MPtw}{\mathcal{M}_{\mathrm{P2}}}

\newcommand{\miii}{\langle III \rangle}

\newcommand{\mxxx}{\langle XXX \rangle}
\newcommand{\mxxy}{\langle XXY \rangle}
\newcommand{\mxyy}{\langle XYY \rangle}
\newcommand{\myxy}{\langle YXY \rangle}
\newcommand{\myyx}{\langle YYX \rangle}
\newcommand{\myyy}{\langle YYY \rangle}
\newcommand{\myxx}{\langle YXX \rangle}
\newcommand{\mxyx}{\langle XYX \rangle}

\newcommand{\ktarget}{\ket{\psi_{\mathrm{t}}}}
\newcommand{\btarget}{\bra{\psi_{\mathrm{t}}}}

\newcommand{\Pset}{\mathbf{P}}
\newcommand{\Psettarget}{\mathbf{P}_{\mathrm{t}}}
\newcommand{\Po}{\overset{\rightarrow}{P_1}}
\newcommand{\Ptw}{\overset{\rightarrow}{P_2}}
\newcommand{\Pth}{\overset{\rightarrow}{P_3}}
\newcommand{\Potw}{\overset{\Rightarrow}{P_{12}}}
\newcommand{\Poth}{\overset{\Rightarrow}{P_{13}}}
\newcommand{\Ptwth}{\overset{\Rightarrow}{P_{23}}}
\newcommand{\Potwth}{\overset{\Rrightarrow}{P_{123}}}

\newcommand{\MHz}{\mathrm{MHz}}
\newcommand{\GHz}{\mathrm{GHz}}
\newcommand{\us}{\mu\mathrm{s}}
\newcommand{\ns}{\mathrm{ns}}

\newcommand{\mV}{\mathrm{mV}}

\newcommand{\EJqmax}{E_{\mathrm{J}q}^{\mathrm{max}}}

\newcommand{\ECq}{E_{\mathrm{C}q}}
\newcommand{\EJq}{E_{\mathrm{J}q}}

\newcommand{\Tone}{T_{1}}
\newcommand{\Ttwostar}{T_{2}^{\ast}}

\newcommand{\wcav}{\omega_{\mathrm{c}}}

\newcommand{\Rxpt}{R_x^{\pi/2}}
\newcommand{\Rxp}{R_x^{\pi\protect\vphantom{/2}}}

\newcommand{\Rypt}{R_y^{\pi/2}}

\begin{document}
\title{Preparation and Measurement of Three-Qubit Entanglement in a Superconducting Circuit}
\author{L.\ DiCarlo}
\affiliation{Departments of Physics and Applied Physics, Yale University, New Haven, CT 06511, USA}
\author{M.\ D.\ Reed}
\affiliation{Departments of Physics and Applied Physics, Yale University, New Haven, CT 06511, USA}
\author{L.\ Sun}
\affiliation{Departments of Physics and Applied Physics, Yale University, New Haven, CT 06511, USA}
\author{B.\ R. \ Johnson}
\affiliation{Departments of Physics and Applied Physics, Yale University, New Haven, CT 06511, USA}
\author{J.\ M. \ Chow}
\affiliation{Departments of Physics and Applied Physics, Yale University, New Haven, CT 06511, USA}
\author{J.\ M. \ Gambetta}
\affiliation{Department of Physics and Astronomy and Institute for Quantum Computing, University of Waterloo, Waterloo, Ontario N2L 3G1, Canada}
\author{L.\ Frunzio}
\affiliation{Departments of Physics and Applied Physics, Yale University, New Haven, CT 06511, USA}
\author{S.\ M. \ Girvin}
\affiliation{Departments of Physics and Applied Physics, Yale University, New Haven, CT 06511, USA}
\author{M.\ H. \ Devoret}
\affiliation{Departments of Physics and Applied Physics, Yale University, New Haven, CT 06511, USA}
\author{R.\ J. \ Schoelkopf}
\affiliation{Departments of Physics and Applied Physics, Yale University, New Haven, CT 06511, USA}
\date{\today}
\maketitle

\textbf{
Traditionally, quantum entanglement has  played a central role in foundational discussions of quantum mechanics.
The measurement of correlations between entangled particles can exhibit results at odds with classical behavior. These discrepancies
increase exponentially with the number of entangled particles~\cite{Mermin90}. When entanglement is extended from just two quantum bits (qubits) to three, the incompatibilities between classical and quantum correlation properties can change from a violation of inequalities~\cite{Bell64} involving statistical averages to sign differences in deterministic observations~\cite{Greenberger89}. With the ample confirmation of quantum mechanical predictions by experiments~\cite{Aspect82,Pan00,Zhao03,Zhao04},  entanglement has evolved from a philosophical conundrum to a key resource for quantum-based technologies, like quantum cryptography and computation~\cite{Nielsen00}. In particular, maximal entanglement of more than two qubits is crucial to the implementation of quantum error correction protocols. While entanglement of up to 3, 5, and 8  qubits has been demonstrated among spins~\cite{Neumann08},  photons~\cite{Zhao04}, and ions~\cite{Haffner05}, respectively, entanglement in engineered solid-state systems has been limited to two qubits~\cite{Steffen06,Leek09, DiCarlo09,Ansmann09,Chow09}. Here, we demonstrate three-qubit entanglement in a superconducting circuit, creating Greenberger-Horne-Zeilinger (GHZ) states with fidelity of $\mathbf{88}\boldsymbol{\%}$, measured with quantum state tomography. Several entanglement witnesses show violation of bi-separable bounds by $\mathbf{830}\boldsymbol{\pm}\mathbf{80}\boldsymbol{\%}$. Our
entangling sequence realizes the first step of basic quantum error correction, namely the encoding of a logical qubit into a manifold of GHZ-like states using a repetition code. The integration of encoding, decoding and error-correcting steps in a feedback loop will be the next milestone for quantum computing with integrated circuits.
}

\begin{figure}[htbp!]
%\psfrag{A}[c][c][0.8]{$\Fth$}
%\psfrag{B}[c][c][0.8]{$\Ftw$}
%\psfrag{C}[c][c][0.8]{$\Fo$}
%\psfrag{D}[c][c][0.8]{$\Fbare$}
%\psfrag{E}[c][c][0.8]{$\Vo$}
%\psfrag{K}[c][c][0.8]{$\Vtw$}
%\psfrag{G}[c][c][0.8]{$\Vth$}
%\psfrag{H}[c][c][0.8]{$\Vf$}
%\psfrag{J}[c][c][0.8]{$\VH$}
\centering
\includegraphics[width=89mm]{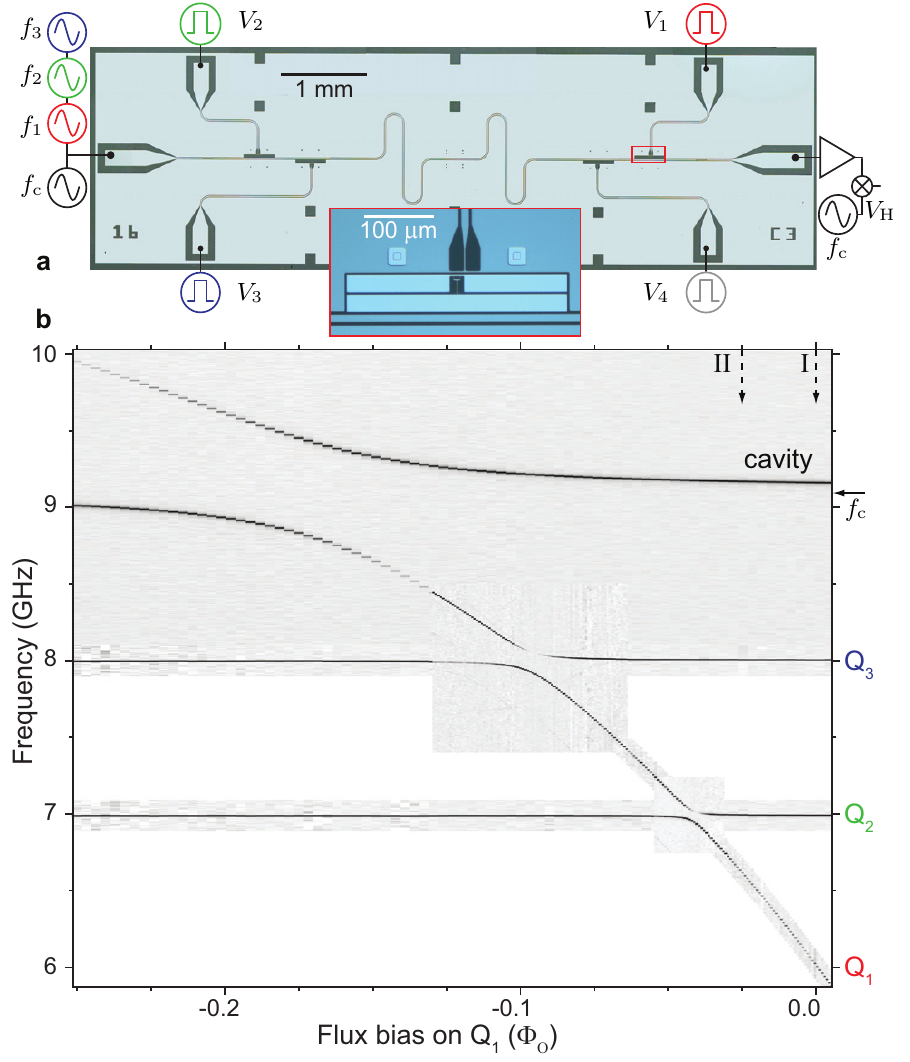}
\caption{{\bfseries Four-qubit cQED processor, and spectroscopic characterization}.
\textbf{a,}~Micrograph of 6-port superconducting device with four transmon qubits [$\Qo$ (inset) to $\Qf$] inside a meandering coplanar waveguide resonator. Local flux-bias lines allow qubit tuning on nanosecond timescales with room-temperature voltages $V_i$. Microwave pulses at qubit transition frequencies $\Fo$, $\Ftw$, and $\Fth$ realize single-qubit $x$- and $y$-rotations in 8~ns. $\Qf$ (operational but unused) is biased at its maximal frequency of $12.27~\GHz$ to minimize its interaction with the qubits employed. Pulsed measurement of cavity homodyne voltage $\VH$ (at the bare cavity frequency $\Fbare=9.070~\GHz$) allows joint qubit readout. A detailed schematic of the measurement setup is shown in Supplementary Fig.~S2.
\textbf{b,}~Grey-scale images of cavity transmission and qubit spectroscopy versus local tuning of $\Qo$ show avoided crossings with $\Qtw$ ($66~\MHz$ splitting), with $\Qth$ ($128~\MHz$ splitting) and with cavity ($615~\MHz$ splitting). Points~I and~II are two of three operating points (Fig.~\ref{fig:fig2} shows III). Single-qubit gates and joint readout are performed at I. A C-Phase gate between $\Qo$ and $\Qtw$ is achieved by flux pulsing to II.
\label{fig:fig1}}
\end{figure}

With  steady improvements in qubit coherence, control, and readout over a decade, superconducting quantum circuits~\cite{Clarke08} have recently
attained two milestones for solid-state two-qubit entanglement. The first is the violation of Bell inequalities without a detection loophole,
realized with phase qubits by minimizing cross-talk between high-fidelity individual qubit readouts~\cite{Ansmann09}. Second is the realization of simple quantum algorithms~\cite{DiCarlo09}, achieved through improved two-qubit gates and coherence in circuit quantum electrodynamics (cQED)~\cite{Wallraff04,Blais04}. The extension of solid-state entanglement from two to several qubits is a new milestone within reach of existing technology. Tripartite interactions between two phase qubits and a resonant cavity in cQED were recently demonstrated~\cite{Altomare10}, suggesting a deterministic but unverified production of  W-class tripartite entanglement~\cite{Acin01}. Here, we demonstrate the extension of conditional-phase gates~\cite{DiCarlo09} (C-Phase) and joint qubit readout~\cite{Filipp09,Chow09} in cQED to the generation and detection of more stringent GHZ-class entanglement between three superconducting charge qubits. Independently, entanglement between three phase qubits has been created and detected at UCSB, and is reported in a simultaneous publication~\cite{Neeley10}.

Our superconducting chip (Fig.~\ref{fig:fig1}a) consists of four  transmon qubits~\cite{Koch07,Schreier08} (labeled $\Qo$ to $\Qf$ counter-clockwise from top right) inside a transmission-line cavity that couples them~\cite{Majer07}, isolates them from the electromagnetic environment~\cite{Houck08}, and allows their joint readout~\cite{Filipp09,Chow09,Reed10}. As in the two-qubit predecessor~\cite{DiCarlo09,Chow09}, qubit control is achieved with a combination of resonant microwave drives realizing single-qubit $x$- and $y$-rotations, and flux pulses individually tuning the qubit transition frequencies on nanosecond timescales. Flux pulses inducing small frequency excursions ($\lesssim 100~\MHz$) realize $z$-rotations. Stronger pulses
($\sim650~\MHz$ excursions) drive specific computational levels into resonance with  non-computational ones (involving second-excited states of $\Qtw$ and $\Qth$) to realize C-Phase gates between nearest neighbors in frequency~\cite{Strauch03}. The readout exploits qubit-state-dependent cavity transmission to gain direct access to multi-qubit correlations, facilitating full tomography of the qubit register and entanglement witnessing. We emphasize that doubling the number of coupled qubits has been achieved without significantly increasing the complexity of circuit design, sample fabrication, or experimental calibration, demonstrating the power of a quantum bus architecture.

\begin{figure}[t!]
%\psfrag{A}[c][c][1.0]{$\pi$}
%\psfrag{t}[c][c][1.0]{$\tau$}
%\psfrag{Q}[c][c][1.0]{$\delta \Vtw$}
%\psfrag{C}[c][c][1.0]{$\tau$}
%\psfrag{X}[c][c][1]{$\Vo$}
%\psfrag{W}[c][c][1]{$\Vtw$}
\centering
\includegraphics[width=89mm]{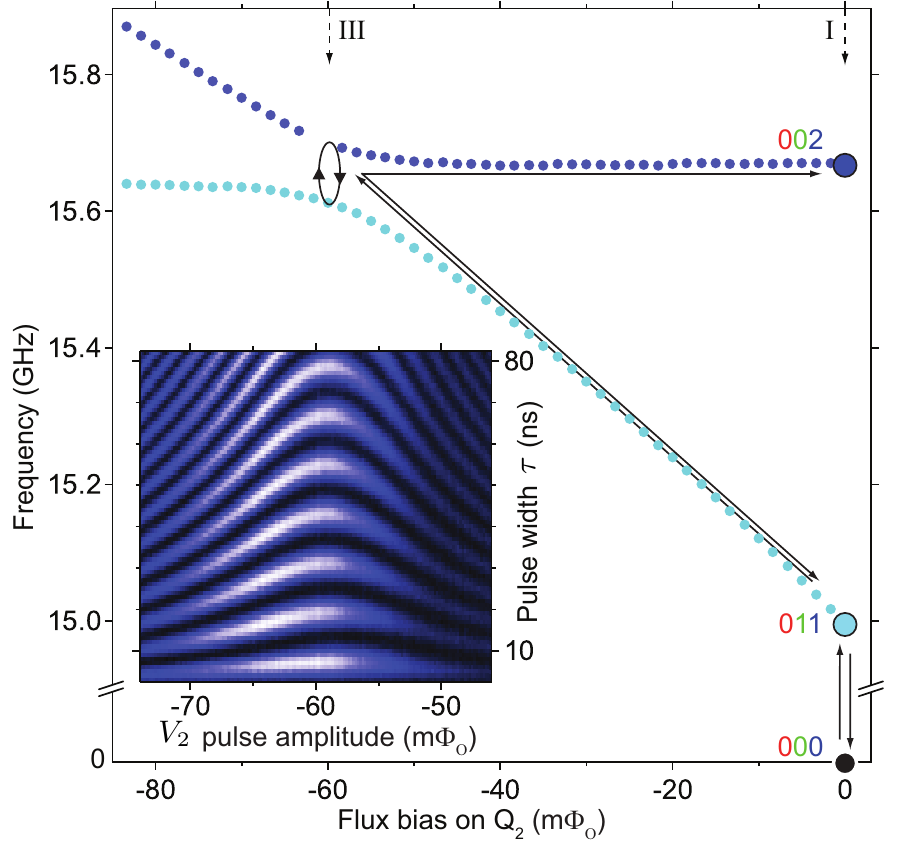}
\caption{{\bfseries Frequency- and time-domain characterization of two-qubit-gate primitive}.
\textbf{Main,}~Two-tone spectroscopy~\cite{Schreier08} of computational level $\ket{011}$
and non-computational level $\ket{002}$  through their avoided crossing ($86~\MHz$ splitting) at point III.
This crossing (and its $\ket{111}\!\leftrightarrow\!\ket{102}$ analog in the three-excitation manifold) is the primitive  for a
C-Phase gate between $\Qtw$ and $\Qth$  (see Ref.~\onlinecite{Strauch03}). The gate is realized with a sudden flux pulse into III. While the pulse is on, the quantum amplitude initially in $\ket{011}$ is coherently exchanged with $\ket{002}$. The pulse is turned off after one full period, at which time all quantum amplitude returns to  $\ket{011}$, but with an additional phase of $\pi$. \textbf{Inset,}~Time-domain characterization of the avoided crossing using the sequence outlined by arrows in main panel. Starting from $\ket{000}$, simultaneous $\pi$ pulses on $\Qtw$ and $\Qth$ populate $\ket{011}$. A $\Vtw$ pulse of duration $\tau$ is next applied. Simultaneous $\pi$-pulses then transfer the final quantum amplitude in $\ket{011}$ to $\ket{000}$ to maximize readout contrast. This characterization gives a calibration of the optimal flux-pulse duration, in this case $12~\ns$.
\label{fig:fig2}}
\end{figure}

\begin{figure*}[ht!]
%\psfrag{A}[c][c][1.0]{$\Rypt$}
%\psfrag{D}[c][c][1.0]{$\Rtomo$}
%\psfrag{P}[c][c][1.0]{$\Po$}
%\psfrag{O}[c][c][1.0]{$\Ptw$}
%\psfrag{R}[c][c][1.0]{$\Pth$}
%\psfrag{S}[c][c][1.0]{$\Potw$}
%\psfrag{T}[c][c][1.0]{$\Poth$}
%\psfrag{U}[c][c][1.0]{$\Ptwth$}
%\psfrag{V}[c][c][1.0]{$\Potwth$}
%\psfrag{E}[c][c][1.0]{$\ket{\psi_{\mathrm{t}}}\!=\!\ket{000}$}
%\psfrag{F}[c][c][1.0]{$\ket{\psi_{\mathrm{t}}}\!=\!\ket{0}\!\otimes\!\left(\ket{00}\!+\!\ket{11}\right)/\sqrt{2}$}
%\psfrag{G}[c][c][1.0]{$\ket{\psi_{\mathrm{t}}}\!=\!\left(\ket{000}\!+\!\ket{111}\right)/\sqrt{2}$}
%\psfrag{J}[c][c][1.0]{$F\!=\!99\%$}
%\psfrag{K}[c][c][1.0]{$F\!=\!94\%$}
%\psfrag{L}[c][c][1.0]{$F\!=\!88\%$}
%\psfrag{I}[c][c][0.7]{$I$}
%\psfrag{X}[c][c][0.7]{$X$}
%\psfrag{Y}[c][c][0.7]{$Y$}
%\psfrag{Z}[c][c][0.7]{$Z$}
\centering
\includegraphics[width=183mm]{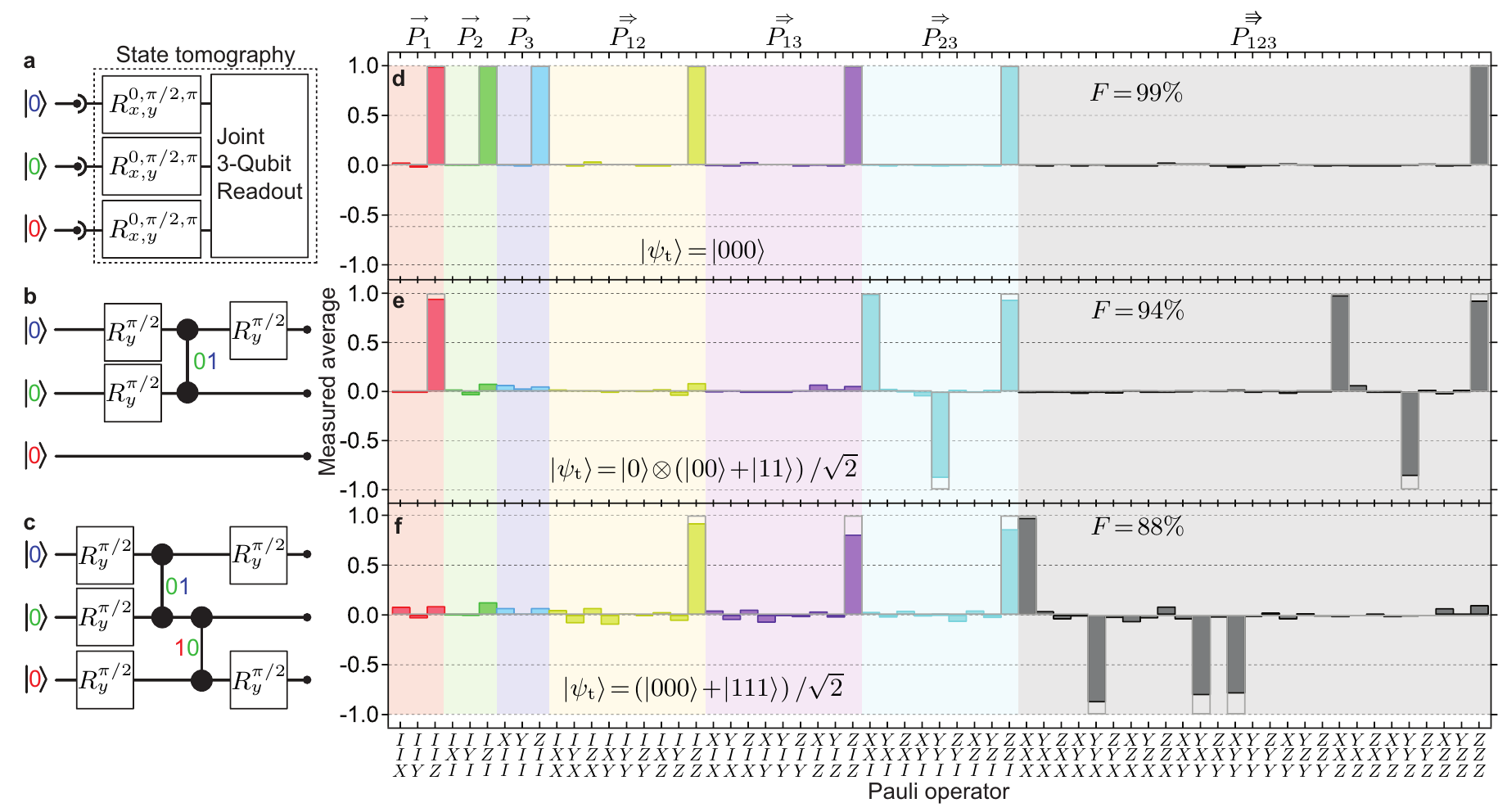}
\caption{{\bfseries Building three-qubit entanglement with two-qubit gates.} \textbf{a},\textbf{b},\textbf{c}, Gate sequences producing states with increasing number of entangled qubits: (\textbf{a}) the ground state (no entanglement), (\textbf{b}) a Bell triplet ($\Qtw$ and $\Qth$ entangled), and (\textbf{c}) the Greenberger-Horne-Zeilinger (GHZ) state (three-qubit entanglement). Vertical lines terminating in solid circles represent C-Phase gates. The coloured two-bit  number next to each indicates the computational basis states that acquire the $\pi$ phase. The state tomography sequence shown in \textbf{a} is also applied in \textbf{b} and \textbf{c}.  \textbf{d},\textbf{e},\textbf{f}, Reconstructed density matrices of the sequence outputs, visualized with a bar chart of the Pauli set $\Pset$. Colors denote the seven subsets of $\Pset$ [Bloch vectors ($\Po$, $\Ptw$, $\Pth$), two-qubit correlations ($\Potw$, $\Poth$, $\Ptwth$) and three-qubit correlations ($\Potwth$)]. The Pauli set $\Psettarget$ of the target state $\ktarget$ is superposed (open bars). The $\Psettarget$ have seven non-zero and full-magnitude bars because they represent stabilizer states~\cite{Nielsen00}. For the GHZ state, they appear exclusively in the correlations, a hallmark of maximal three-qubit entanglement. The experimental $\Pset$ closely match $\Psettarget$ in the three cases, with fidelities $F=\btarget \rho \ktarget=\Pset\cdot\Psettarget/8$ of $0.99$, $0.94$, and $0.88$. By exceeding $50\%$, the fidelity to the GHZ state witnesses genuine three-qubit entanglement (3QE). By exceeding $75\%$, it witnesses the stringent GHZ-class of 3QE~\cite{Acin01}. More traditional cityscape visualizations of the density matrices are shown in Supplementary Fig.~S4.
\label{fig:fig3}}
\end{figure*}

The spectrum of single excitations of the three employed qubits ($\Qo$ to  $\Qth$) and the cavity reveals key features of the generalized Tavis-Cummings Hamiltonian and allows extraction of its parameters (see Methods).  Spectroscopy as a function of local flux tuning of $\Qo$ (Fig.~\ref{fig:fig1}b) shows exactly three avoided crossings: $J$-crossings~\cite{Majer07} with $\Qtw$ and with $\Qth$, and the vacuum-Rabi splitting~\cite{Wallraff04} with the cavity near its bare frequency $\Fbare=9.070~\GHz$. To the resolution of all spectroscopy performed, the spectrum is free of spurious avoided crossings. This is a critical requirement for pulsed excursions of qubit transition frequencies. We choose point~I [$(\Fo,\Ftw,\Fth)=(6.000,7.000,8.000)~\GHz \pm 2~\MHz$] for all single-qubit rotations and for readout. Here, the qubits are sufficiently detuned from their nearest neighbors in frequency to make their interaction small, yet close enough to the cavity to reach the strong-dispersive regime of cQED~\cite{Gambetta06,Schuster07}.

Two-qubit C-Phase gates are the  workhorses that generate entanglement in the qubit register.  We realize C-Phase gates by direct extension of the protocol proposed for phase qubits in Ref.~\onlinecite{Strauch03}, wherein a full coherent oscillation between computational and non-computational states yields a two-qubit phase of $\pi$.  The primitive interaction for C-Phase between $\Qtw$ and $\Qth$ is shown with two-tone spectroscopy and time-domain data in Fig.~\ref{fig:fig2}. (See Supplementary Fig.~S1 for a similar characterization of the primitive for C-Phase between $\Qo$ and $\Qtw$). At point III, the computational level $\ket{011}$ becomes resonant with the non-computational level $\ket{002}$ ($\ket{abc}$ denotes excitation level $a$ on $\Qo$, $b$ on $\Qtw$ and $c$ on $\Qth$). The cavity-mediated interaction between these levels produces an avoided crossing of $86~\MHz$. An analogous  avoided crossing takes places simultaneously in the three-excitation manifold, between  $\ket{111}$ and $\ket{102}$. A coherent oscillation between the computational and the non-computational levels is started by pulsing non-adiabatically into point III. A full oscillation is completed in 12~ns (Fig.~\ref{fig:fig2} inset), returning all the quantum amplitude to the computational level but with an additional phase of $\pi$. The two-qubit gate time is nearly half that of our previous implementation, which used the avoided crossing adiabatically~\cite{DiCarlo09}. To complete the C-Phase gate, the single-qubit dynamical phase acquired by $\Qtw$ during the flux pulse (and also by $\Qo$ and $\Qth$ through residual flux cross-talk) is canceled using a $z$-rotation (see Supplementary Fig.~S3).

To detect the entanglement produced with C-Phase gates, we employ a high-fidelity three-qubit joint readout presented in a parallel publication~\cite{Reed10}. This readout allows an approximately ten-fold increase in single-shot fidelity over the previous two-qubit joint readout~\cite{Filipp09,DiCarlo09,Chow09} without requiring any additional hardware or design modification. Readout was previously performed by a pulsed measurement of $\VH$ in linear response ($\sim1$ photon in cavity). Here, we drive the cavity at $\Fbare$ with 50,000 times larger incident power,  well past the onset ($\sim10$ photons) of the non-linearity that the cavity inherits via dispersive coupling to the qubits. Turning on this strong drive can make the cavity excite, conditioned on the three-qubit state, into a high-transmission state where it regains linearity. We adjust the incident power so that the cavity excites for all register states except $\ket{000}$. If this selectivity were perfect, the measurement would be projective on $\ket{000}$, making the ensemble average $\mVH \propto\sum_{A,B,C\in\{I,Z\}}\langle A^{(1)} B^{(2)} C^{(3)} \rangle$. Here, the $A^{(i)}$ represent Pauli operators~\cite{Nielsen00} acting on $Q_i$ (henceforth, the order of operators is respected and superscripts are removed for notational simplicity). False positives and negatives  introduce  weighting coefficients $\beta_{ABC}$. The calibrated values (listed in Methods) demonstrate a high sensitivity of the single measurement channel to two- and three-qubit correlations.

We use this sensitivity to qubit correlations to perform state tomography of the  register. To reconstruct the three-qubit density matrix $\rho$, we find the coefficients of its expansion in the Pauli operator basis:
\begin{equation}
\rho = \frac{1}{8}\sum_{A,B,C \in \{I,X,Y,Z\}}\langle ABC \rangle ABC,
\end{equation}
where $\miii=1$. This is achieved by pre-pending sets of single-qubit rotations to the readout pulse. The rotations consist of all combinations of $I$, $\Rxp$, $\Rxpt$, and $\Rypt$ on the three qubits (except for $\Rxp \otimes \Rxp \otimes \Rxp$). Respectively, these rotations on $Q_i$ transform the $\mVH$ expression according to $Z^{(i)}\rightarrow Z^{(i)}$, $-Z^{(i)}$, $Y^{(i)}$, and $-X^{(i)}$. An ensemble of $10^5$ state preparations and single-shot measurements are made for each set, mitigating uncertainty due to projection noise to less than $1\%$. The non-trivial correlation coefficients in Eq.~(1) are then obtained from the 63 distinct $\mVH$ by matrix inversion.

With  fast C-Phase gates and high fidelity readout in place, we now demonstrate generation and detection of  multi-qubit entanglement.
Gate sequences generating two- and three-qubit entanglement are shown in Fig.~\ref{fig:fig3}.
A simple sequence~\cite{DiCarlo09} using one C-Phase transforms the ground state $\ket{000}$ (an unentangled, or separable state) into  a Bell triplet $\ket{0} \otimes \left(\ket{00}+\ket{11}\right)/\sqrt{2}$ with $\Qtw$ and $\Qth$ maximally entangled (Fig.~\ref{fig:fig3}b). Mirroring this sequence so that $\Qo$ undergoes the same operations as $\Qth$ (Fig.~\ref{fig:fig3}c)  produces the GHZ state, $\GHZ=\left(\ket{000}+\ket{111}\right)/\sqrt{2}$, a maximally-entangled state of three qubits.   We have implemented these sequences (see Supplementary Fig.~S3 for the actual microwave and flux pulses realizing the GHZ sequence) and performed tomography of their outputs. We visualize the reconstructed $\rho$ in each case using the Pauli set $\Pset$, consisting of the expectation values of the non-trivial Pauli operators.  In Figs.~\ref{fig:fig3}d-f, we subdivide $\Pset$ into seven subsets distinguished by color: three single-qubit Bloch vectors $\Po$ (red), $\Ptw$ (green), and $\Pth$ (blue); two-qubit correlations $\Potw$ (orange), $\Poth$ (purple), and $\Ptwth$ (cyan); and three-qubit correlations $\Potwth$ (grey). The experimental $\Pset$ in  Figs.~\ref{fig:fig3}d-f closely match the delineated Pauli set $\Psettarget$ of the targeted ground state, the Bell triplet and the GHZ state, respectively. We quantify this similarity using fidelity to the target state $\ktarget$, $F=\btarget \rho \ktarget=\Pset\cdot\Psettarget/8$, finding $F=99$, 94 and 88\%, respectively.

To make definitive statements about the presence of genuine three-qubit entanglement (3QE)  in  Fig.~\ref{fig:fig3}f, we make use of
fidelity to GHZ states as an entanglement witness~\cite{Acin01}. The maximal fidelity of any bi-separable state to a GHZ state is $50\%$. Any greater fidelity thus witnesses 3QE. Fidelity can even witness the more restrictive GHZ-class within 3QE, since W-class states satisfy $F\leq 75\%$. The 88\% fidelity to $\GHZ$ of the Pauli set in Fig.~\ref{fig:fig3}f constitutes the first demonstration of GHZ-type entanglement between three engineered solid-state qubits.

\begin{figure}[htbp!]
%\psfrag{A}[c][c][1.0]{$\Rypt$}
%\psfrag{B}[c][c][1.0]{$\Rtomo$}
%\psfrag{C}[c][c][1.0]{$R_{\hat{n}(\!\phi\!)}^{\pi/2}$}
%\psfrag{D}[c][c][1.0]{$\MSo$}
%\psfrag{E}[c][c][1.0]{$\MStw$}
%\psfrag{G}[c][c][0.95]{$\MPo$}
%\psfrag{H}[c][c][0.95]{$\MPtw$}
%\psfrag{p}[c][c][1]{$\phi$}
%\psfrag{n}[c][c][1]{$\widehat{n}$}
\centering
\includegraphics[width=89mm]{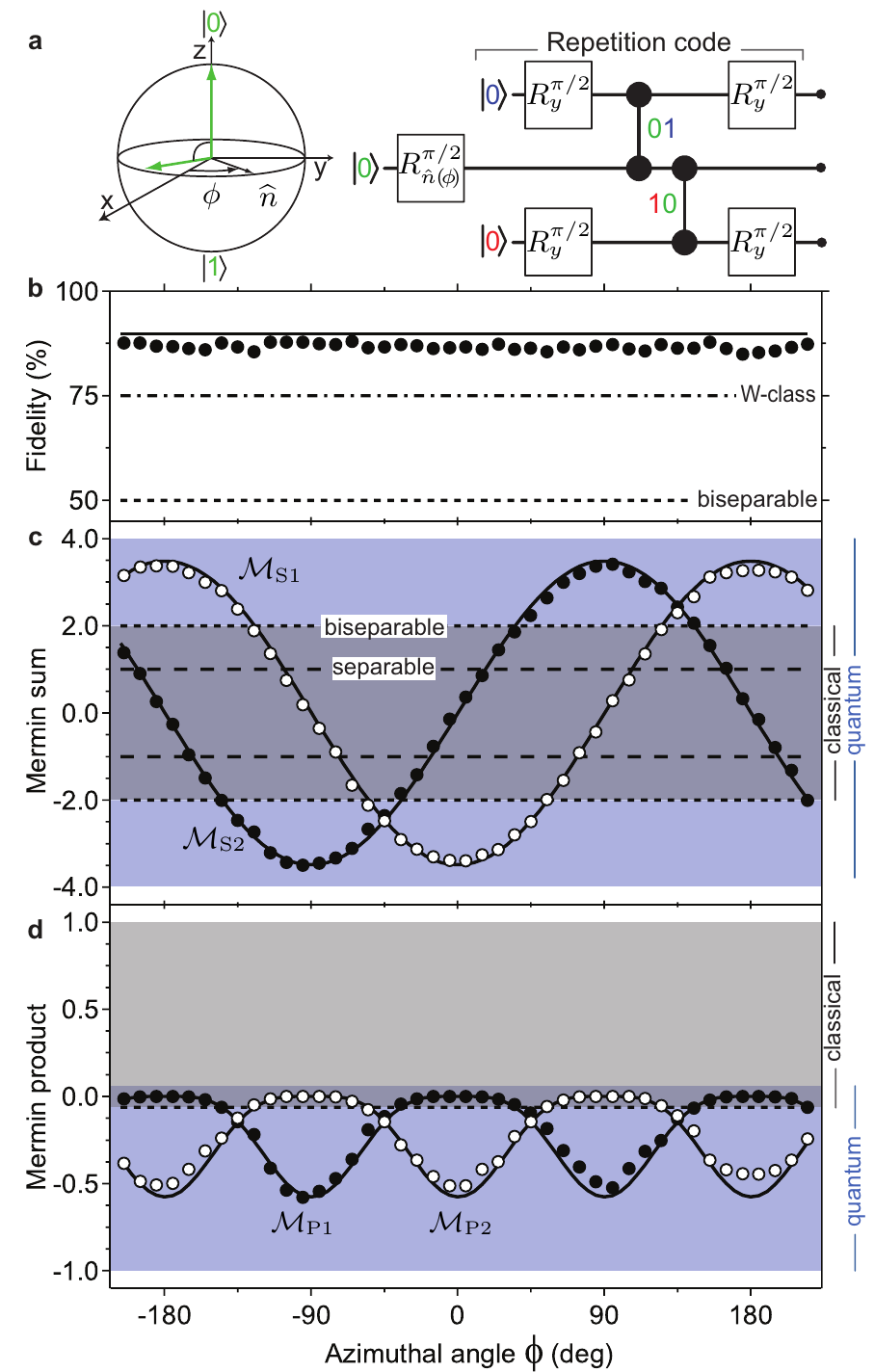}
\caption{{\bfseries Witnessing of three-qubit entanglement using fidelity and Mermin inequalities}.
\textbf{a,}~Gate sequence mapping superpositions $\left(\ket{0}+e^{i(\phi-\pi/2)}\ket{1}\right)/\sqrt{2}$ of $\Qtw$ into GHZ states $\GHZphi=\left(\ket{000}+e^{i(\phi-\pi/2)}\ket{111}\right)/\sqrt{2}$  using a repetition code. \textbf{b,}~Fidelity $F=\bra{\mathrm{GHZ}_{\phi}}\rho\ket{\mathrm{GHZ}_{\phi}}$
as a function of azimuthal angle $\phi$ of initial $\pi/2$-rotation on $\Qtw$, averaging $87\%$. Bi-separable ($F\leq50\%$) and W-class 3QE ($F\leq75\%$) bounds are amply exceeded, witnessing stringent GHZ-class 3QE. \textbf{c,} Evolution of Mermin sums $\MSo\!\!=\!\mxxx\!-\!\myyx\!-\!\myxy\!-\!\mxyy$ and $\MStw=-\myyy+\mxxy+\mxyx+\myxx$. Separable ($|\MS|\leq1$) and bi-separable ($|\MS|\leq2$) bounds are broken by at least one sum at each $\phi$, detecting 3QE. $|\MS|\leq2$ (gray shading) is also the LHV bound. The extremal measured value $3.4\pm0.1$ largely exceeds it. The oscillations are fully within the range allowed by quantum mechanics assuming three qubits $|\MS|\leq4$ (blue shading). \textbf{d,} Evolution of Mermin products $\MPo\!=\!\mxxx\myyx\myxy\mxyy$ and $\MPtw\!=\!\myyy\mxxy\mxyx\myxx$.
The negative bi-separable (also LHV) bound is  $\MP \geq -1/16$.  The minimum value measured of $-0.52\pm0.05$ detects 3QE with a violation of $830\pm80\%$. Solid curves in the three panels correspond to a master equation simulation that includes qubit relaxation during the pulse sequence.
\label{fig:fig4}}
\end{figure}

The production of multi-qubit entanglement is a necessary first step toward quantum error correction.  In fact, the simple sequence using two C-Phase gates (Fig.~\ref{fig:fig3}c) performed the encoding step of the simplest error correction protocol, the bit-flip code~\cite{Nielsen00}. Generally, this encoding maps a logical qubit state $\alpha \ket{0} + \beta \ket{1}$ onto the state  $\alpha \ket{000} + \beta \ket{111}$ of three physical qubits. In
Fig.~\ref{fig:fig3}c,  the encoding was performed specifically for the state $\left(\ket{0} +\ket{1}\right)/\sqrt{2}$ in $\Qtw$. We have applied this  repetition code to other maximal superpositions of $\Qtw$ by varying the azimuthal angle $\phi$ of its initial $\pi/2$ rotation (Fig.~\ref{fig:fig4}a). At each $\phi$, the code targets a GHZ state $\GHZphi=\left(\ket{000}+e^{i(\phi-\pi/2)}\ket{111}\right)/\sqrt{2}$. The fidelity to $\GHZphi$ is $87\pm1\%$ throughout (Fig.~\ref{fig:fig4}b). A master equation simulation suggests that this uniform fidelity is largely limited by qubit relaxation during the $81~\ns$ pulse sequence. The measured fidelity witnesses GHZ-class 3QE at every $\phi$.

It is possible to detect three-qubit entanglement with linear witnesses which can be computed using fewer elements of the Pauli set than the fidelity to a GHZ state. For example, the Mermin sums~\cite{Mermin90} $\MSo=\mxxx-\mxyy-\myxy-\mxyy$ and $\MStw=-\myyy+\myxx+\mxyx+\myxx$
satisfy $|\MS|\leq1$ and $|\MS|\leq2$ for all separable~\cite{Roy05} and bi-separable~\cite{Toth05} states, respectively. Figure~\ref{fig:fig4}c shows that at least one of these sums detects 3QE at each $\phi$. Note that $|\MS|\leq2$ is also a local-hidden-variable (LHV) bound~\cite{Mermin90}. While the maximal absolute value measured, $3.4\pm0.1$, exceeds this bound by 14 standard deviations, the presence of locality and detection loopholes in our system precludes the refutation of local realism.

One drawback of the Mermin sums as witnessses of 3QE is that the bi-separable (and LHV) range overlaps significantly with the quantum range for three qubits, $|\MS|\leq4$. Non-linear entanglement witnesses can compress the bi-separable bounds relative to the quantum bounds, effectively magnifying non-trivial three-qubit correlations. We have investigated the Mermin products $\MPo=\mxxx\mxyy\myxy\mxyy$ and $\MPtw=\myyy\myxx\mxyx\myxx$, finding various bounds numerically. Separable and bi-separable states satisfy $0\leq\MP\leq1/64$ and $-1/16\leq\MP\leq1/64$, respectively. Their range is only a small fraction of the quantum range for three qubits, $-1\leq\MP\leq1/16$, as advertised. The LHV range for Mermin products is $-1/16\leq\MP\leq 1$. Note that while the LHV range for Mermin sums is fully inside the range allowed by quantum mechanics, the two ranges largely separate for products, leaving only a narrow region of compatibility $|\MP|\leq1/16$.  The measured Mermin products (Fig.~\ref{fig:fig4}d)  reach a minimum value $-0.52\pm0.05$, exceeding the negative bi-separable (also LHV) bound by $830\pm80\%$. The experimental $\MP$ fall largely outside the compatibility region, and fully within the quantum bounds.

We have applied conditional-phase  gates and joint readout in cQED to produce and detect GHZ-class entanglement between three superconducting qubits.
Extending solid-state entanglement beyond two qubits has not required significantly more complex circuit design, fabrication or calibration, and worked the first time. We have generated GHZ states with fidelity approaching $90\%$ and detected their entanglement using quantum state tomography as well as various linear and nonlinear entanglement witnesses requiring fewer measurements. Finally, we have realized the first step of basic quantum error correction, namely the encoding of one logical qubit in an entangled state of three physical qubits using a repetition code. Future research will focus on the realization of an error-syndrome detecting circuit to allow closing the feedback loop needed for error correction.

\bigskip\noindent
\textbf{Acknowledgements}
We thank M.~Brink for experimental contributions,
and L.~S.~Bishop, E.~Ginossar, D.~I.~Schuster, and C.~Rigetti for discussions.
We acknowledge support from LPS/NSA under ARO Contract No.\ W911NF-05-1-0365,
from IARPA under ARO Contract No.\ W911NF-09-1-0369,
and from the NSF under Grants No.\ DMR-0653377 and No.\ DMR-0603369.
Additional support provided by CNR-Istituto di Cibernetica, Pozzuoli, Italy (LF),
and by CIFAR, MITACS, MRI, and NSERC (JMG). All statements of fact, opinion or conclusions
contained herein are those of the authors and should not be construed as
representing the official views or policies of the U.S. Government. 

\section{Methods}
\noindent
\textbf{Hamiltonian parameters}
The Tavis-Cummings Hamiltonian generalized to transmons is
\begin{eqnarray}
\label{eq:Ham}
H &=&\hbar\wcav a^{\dag}a + \nonumber\\
  & & \hbar \sum_{q=1}^4 \Bigl( \sum_{j=0}^{N} \omega_{0j}^{(q)} \lvert j\rangle_q \langle j\rvert_q + (a+a^{\dag}) \!\!\sum_{j,k=0}^{N} g_{jk}^{(q)}\lvert j\rangle_q\langle k\rvert_q \Bigr). \nonumber
\end{eqnarray}
Here, $\hbar$ is the reduced Planck constant, $\wcav$ is the bare cavity frequency, $\omega^{(q)}_{0j}$ is the transition frequency for qubit $q$ from ground to excited state $j$, and $g_{jk}^{(q)}=g_q n_{jk}$, with  $g_q$ a bare qubit-cavity coupling and $n_{jk}$ a level-dependent coupling matrix element. Both $\omega^{(q)}_{0j}$ and $n_{jk}$ are functions~\cite{Koch07} of qubit charging energy $\ECq$  and  Josephson energy $\EJq$. The flux control enters through $\EJq=\EJqmax\lvert\cos(\pi\Phi_q/\Phi_0)\rvert$, with  $\Phi_q$ the flux through the transmon SQUID loop, and a linear flux-voltage relation $\Phi_q=\sum_{i=1}^4\alpha_{qi} V_i + \Phi_{q,0}$ that includes cross-talk and offsets ($\Phi_0$ is the flux quantum). Cross-talk (up to $\sim40\%$) resulting from return currents on the ground plane is corrected by orthogonalization. The above parameters are  constrained by the spectroscopy and transmission data shown (Figs.~\ref{fig:fig1}b, \ref{fig:fig2}, and S2) and similar data (not shown) obtained as a function of local flux bias on $\Qtw$ to $\Qf$.  Fitting spectra obtained by numerical diagonalization of the Hamiltonian (truncated to  $N=4$ qubit levels and 4 cavity photons) to these data gives
$\wcav/2\pi=9.070~\GHz$,
$\EJqmax/h=\{42,29,47,57\}~\GHz$ (from $\Qo$ to $\Qf$), $g/2\pi \approx 220~\MHz$, and $\ECq/h\approx 330~\MHz$ .

\smallskip\noindent
\textbf{Coherence times}
Relaxation ($\Tone$) and dephasing ($\Ttwostar$) times of $\Qo$ to $\Qth$ were measured using standard sliding $\pi$-pulse and Ramsey experiments, respectively. At point I, $\Tone=(1.2,1.0,0.6)~\us$ ($\Qo$ to $\Qth$) are consistent with relaxation due to the Purcell effect~\cite{Houck08} and non-radiative loss with quality factor $\sim 55,000$. $\Ttwostar=(0.3,0.6,0.5)~\us$ are consistent with  $1/f$ flux noise of $\sim10^{-5}~\Phi_0/\sqrt{\mathrm{Hz}}$ at 1~Hz. The cavity linewidth is $\kappa/2\pi = 2.4\,\MHz$.

\smallskip\noindent
\textbf{Joint readout}
The weighting coefficients $\beta$ in the measurement operator are calibrated in every tomography run by applying joint readout to the eight computational basis states, prepared using combinations of $\pi$ pulses. For example, the ensemble-averaged joint readout of $\ket{101}$ gives
$\langle \VH \rangle = \beta_{III}-\beta_{ZII}+\beta_{IZI}-\beta_{IIZ}-\beta_{ZZI}+\beta_{ZIZ}-\beta_{IZZ}+\beta_{ZZZ}$. The calibration measurements provide eight linearly-independent combinations of the coefficients, and the coefficients are obtained by matrix inversion. The typical values $\{\beta_{ZII},\beta_{IZI},\beta_{IIZ},\beta_{ZZI},\beta_{ZIZ},\beta_{IZZ},\beta_{ZZZ}\}
=\{2.2,3.1,3.2,1.9,2.0,2.9,1.7\}~\mV$ reveal a high sensitivity to two- and three-qubit correlations.

\bibliography{References_GHZ}

\begin{thebibliography}{10}
\expandafter\ifx\csname url\endcsname\relax
  \def\url#1{\texttt{#1}}\fi
\expandafter\ifx\csname urlprefix\endcsname\relax\def\urlprefix{URL }\fi
\providecommand{\bibinfo}[2]{#2}
\providecommand{\eprint}[2][]{\url{#2}}

\bibitem{Mermin90}
\bibinfo{author}{Mermin, N.~D.}
\newblock \bibinfo{title}{Extreme quantum entanglement in a superposition of
  macroscopically distinct states}.
\newblock \emph{\bibinfo{journal}{Phys. Rev. Lett.}}
  \textbf{\bibinfo{volume}{65}}, \bibinfo{pages}{1838--1840}
  (\bibinfo{year}{1990}).

\bibitem{Bell64}
\bibinfo{author}{Bell, J.~S.}
\newblock \bibinfo{title}{On the {E}instein-{P}odolsky-{R}osen paradox}.
\newblock \emph{\bibinfo{journal}{Physics}} \textbf{\bibinfo{volume}{1}},
  \bibinfo{pages}{195--200} (\bibinfo{year}{1964}).

\bibitem{Greenberger89}
\bibinfo{author}{Greenberger, D.~M.}, \bibinfo{author}{Horne, M.~A.} \&
  \bibinfo{author}{Zeilinger, A.}
\newblock \bibinfo{title}{Going beyond {B}ell's theorem}.
\newblock In \bibinfo{editor}{Kafatos, M.} (ed.)
  \emph{\bibinfo{booktitle}{{B}ell's theorem, quantum theory and conceptions of
  the universe}} (\bibinfo{publisher}{Kluwer Academic},
  \bibinfo{address}{Dordrecht}, \bibinfo{year}{1989}).

\bibitem{Aspect82}
\bibinfo{author}{Aspect, A.}, \bibinfo{author}{Dalibard, J.} \&
  \bibinfo{author}{Roger, G.}
\newblock \bibinfo{title}{Experimental test of {B}ell's inequalities using
  time-varying analyzers}.
\newblock \emph{\bibinfo{journal}{Phys. Rev. Lett.}}
  \textbf{\bibinfo{volume}{49}}, \bibinfo{pages}{1804--1807}
  (\bibinfo{year}{1982}).

\bibitem{Pan00}
\bibinfo{author}{Pan, J.-W.}, \bibinfo{author}{Bouwmeester, D.},
  \bibinfo{author}{Daniell, M.}, \bibinfo{author}{Weinfurter, H.} \&
  \bibinfo{author}{Zeilinger, A.}
\newblock \bibinfo{title}{Experimental test of quantum nonlocality in
  three-photon {G}reenberger-{H}orne-{Z}eilinger entanglement}.
\newblock \emph{\bibinfo{journal}{Nature}} \textbf{\bibinfo{volume}{403}},
  \bibinfo{pages}{515--519} (\bibinfo{year}{2000}).

\bibitem{Zhao03}
\bibinfo{author}{Zhao, Z.} \emph{et~al.}
\newblock \bibinfo{title}{Experimental violation of local realism by
  four-photon {G}reenberger-{H}orne-{Z}eilinger entanglement}.
\newblock \emph{\bibinfo{journal}{Phys. Rev. Lett.}}
  \textbf{\bibinfo{volume}{91}}, \bibinfo{pages}{180401}
  (\bibinfo{year}{2003}).

\bibitem{Zhao04}
\bibinfo{author}{Zhao, Z.} \emph{et~al.}
\newblock \bibinfo{title}{Experimental demonstration of five-photon
  entanglement and open-destination teleportation}.
\newblock \emph{\bibinfo{journal}{Nature}} \textbf{\bibinfo{volume}{430}},
  \bibinfo{pages}{54--58} (\bibinfo{year}{2004}).

\bibitem{Nielsen00}
\bibinfo{author}{Nielsen, M.~A.} \& \bibinfo{author}{Chuang, I.~L.}
\newblock \emph{\bibinfo{title}{Quantum Computation and Quantum Information}}
  (\bibinfo{publisher}{Cambridge University Press},
  \bibinfo{address}{Cambridge}, \bibinfo{year}{2000}).

\bibitem{Neumann08}
\bibinfo{author}{Neumann, P.} \emph{et~al.}
\newblock \bibinfo{title}{Multipartite entanglement among single spins in
  diamond}.
\newblock \emph{\bibinfo{journal}{Science}} \textbf{\bibinfo{volume}{320}},
  \bibinfo{pages}{1326--1329} (\bibinfo{year}{2008}).

\bibitem{Haffner05}
\bibinfo{author}{H\"{a}ffner, H.} \emph{et~al.}
\newblock \bibinfo{title}{Scalable multiparticle entanglement of trapped ions}.
\newblock \emph{\bibinfo{journal}{Nature}} \textbf{\bibinfo{volume}{438}},
  \bibinfo{pages}{643--646} (\bibinfo{year}{2005}).

\bibitem{Steffen06}
\bibinfo{author}{Steffen, M.} \emph{et~al.}
\newblock \bibinfo{title}{Measurement of the entanglement of two
  superconducting qubits via state tomography}.
\newblock \emph{\bibinfo{journal}{Science}} \textbf{\bibinfo{volume}{313}},
  \bibinfo{pages}{1423--1425} (\bibinfo{year}{2006}).

\bibitem{Leek09}
\bibinfo{author}{Leek, P.~J.} \emph{et~al.}
\newblock \bibinfo{title}{Using sideband transitions for two-qubit operations
  in superconducting circuits}.
\newblock \emph{\bibinfo{journal}{Phys. Rev. B}} \textbf{\bibinfo{volume}{79}},
  \bibinfo{pages}{180511(R)} (\bibinfo{year}{2009}).

\bibitem{DiCarlo09}
\bibinfo{author}{Di{C}arlo, L.} \emph{et~al.}
\newblock \bibinfo{title}{Demonstration of two-qubit algorithms with a
  superconducting quantum processor}.
\newblock \emph{\bibinfo{journal}{Nature}} \textbf{\bibinfo{volume}{460}},
  \bibinfo{pages}{240--244} (\bibinfo{year}{2009}).

\bibitem{Ansmann09}
\bibinfo{author}{Ansmann, M.} \emph{et~al.}
\newblock \bibinfo{title}{Violation of {B}ell's inequality in {J}osephson phase
  qubits}.
\newblock \emph{\bibinfo{journal}{Nature}} \textbf{\bibinfo{volume}{461}},
  \bibinfo{pages}{504--506} (\bibinfo{year}{2009}).

\bibitem{Chow09}
\bibinfo{author}{Chow, J.~M.} \emph{et~al.}
\newblock \bibinfo{title}{Entanglement metrology using a joint readout of
  superconducting qubits}.
\newblock \bibinfo{howpublished}{arXiv:0908.1955}.

\bibitem{Clarke08}
\bibinfo{author}{Clarke, J.} \& \bibinfo{author}{Wilhelm, F.~K.}
\newblock \bibinfo{title}{Superconducting quantum bits}.
\newblock \emph{\bibinfo{journal}{Nature}} \textbf{\bibinfo{volume}{453}},
  \bibinfo{pages}{1031--1042} (\bibinfo{year}{2008}).

\bibitem{Wallraff04}
\bibinfo{author}{Wallraff, A.} \emph{et~al.}
\newblock \bibinfo{title}{Strong coupling of a single photon to a
  superconducting qubit using circuit quantum electrodynamics}.
\newblock \emph{\bibinfo{journal}{Nature}} \textbf{\bibinfo{volume}{431}},
  \bibinfo{pages}{162--167} (\bibinfo{year}{2004}).

\bibitem{Blais04}
\bibinfo{author}{Blais, A.}, \bibinfo{author}{Huang, R.-S.},
  \bibinfo{author}{Wallraff, A.}, \bibinfo{author}{Girvin, S.~M.} \&
  \bibinfo{author}{Schoelkopf, R.~J.}
\newblock \bibinfo{title}{Cavity quantum electrodynamics for superconducting
  electrical circuits: An architecture for quantum computation}.
\newblock \emph{\bibinfo{journal}{Phys. Rev. A}} \textbf{\bibinfo{volume}{69}},
  \bibinfo{pages}{062320} (\bibinfo{year}{2004}).

\bibitem{Altomare10}
\bibinfo{author}{Altomare, F.} \emph{et~al.}
\newblock \bibinfo{title}{Tripartite interactions between two phase qubits and
  a resonant cavity}.
\newblock \bibinfo{howpublished}{arXiv:1004.0026}.

\bibitem{Acin01}
\bibinfo{author}{Ac\'\i{}n, A.}, \bibinfo{author}{Bru\ss{}, D.},
  \bibinfo{author}{Lewenstein, M.} \& \bibinfo{author}{Sanpera, A.}
\newblock \bibinfo{title}{Classification of mixed three-qubit states}.
\newblock \emph{\bibinfo{journal}{Phys. Rev. Lett.}}
  \textbf{\bibinfo{volume}{87}}, \bibinfo{pages}{040401}
  (\bibinfo{year}{2001}).

\bibitem{Filipp09}
\bibinfo{author}{Filipp, S.} \emph{et~al.}
\newblock \bibinfo{title}{Two-qubit state tomography using a joint dispersive
  readout}.
\newblock \emph{\bibinfo{journal}{Phys. Rev. Lett.}}
  \textbf{\bibinfo{volume}{102}}, \bibinfo{pages}{200402}
  (\bibinfo{year}{2009}).

\bibitem{Neeley10}
\bibinfo{author}{Neeley, M.} \emph{et~al.}
\newblock \bibinfo{title}{Generation of three-qubit entangled states using
  superconducting phase qubits}.
\newblock \emph{\bibinfo{journal}{Submitted}}  (\bibinfo{year}{2010}).

\bibitem{Koch07}
\bibinfo{author}{Koch, J.} \emph{et~al.}
\newblock \bibinfo{title}{Charge-insensitive qubit design derived from the
  {C}ooper pair box}.
\newblock \emph{\bibinfo{journal}{Phys. Rev. A}} \textbf{\bibinfo{volume}{76}},
  \bibinfo{pages}{042319} (\bibinfo{year}{2007}).

\bibitem{Schreier08}
\bibinfo{author}{Schreier, J.~A.} \emph{et~al.}
\newblock \bibinfo{title}{Suppressing charge noise decoherence in
  superconducting charge qubits}.
\newblock \emph{\bibinfo{journal}{Phys. Rev. B}} \textbf{\bibinfo{volume}{77}},
  \bibinfo{pages}{180502(R)} (\bibinfo{year}{2008}).

\bibitem{Majer07}
\bibinfo{author}{Majer, J.} \emph{et~al.}
\newblock \bibinfo{title}{Coupling superconducting qubits via a cavity bus}.
\newblock \emph{\bibinfo{journal}{Nature}} \textbf{\bibinfo{volume}{449}},
  \bibinfo{pages}{443--447} (\bibinfo{year}{2007}).

\bibitem{Houck08}
\bibinfo{author}{Houck, A.~A.} \emph{et~al.}
\newblock \bibinfo{title}{Controlling the spontaneous emission of a
  superconducting transmon qubit}.
\newblock \emph{\bibinfo{journal}{Phys. Rev. Lett.}}
  \textbf{\bibinfo{volume}{101}}, \bibinfo{pages}{080502}
  (\bibinfo{year}{2008}).

\bibitem{Reed10}
\bibinfo{author}{Reed, M.~D.} \emph{et~al.}
\newblock \bibinfo{title}{High fidelity readout in circuit quantum
  electrodynamics using the {J}aynes-{C}ummings nonlinearity}.
\newblock \emph{\bibinfo{journal}{Submitted}}  (\bibinfo{year}{2010}).

\bibitem{Strauch03}
\bibinfo{author}{Strauch, F.~W.} \emph{et~al.}
\newblock \bibinfo{title}{Quantum logic gates for coupled superconducting phase
  qubits}.
\newblock \emph{\bibinfo{journal}{Phys. Rev. Lett.}}
  \textbf{\bibinfo{volume}{91}}, \bibinfo{pages}{167005}
  (\bibinfo{year}{2003}).

\bibitem{Gambetta06}
\bibinfo{author}{Gambetta, J.} \emph{et~al.}
\newblock \bibinfo{title}{Qubit-photon interactions in a cavity:
  Measurement-induced dephasing and number splitting}.
\newblock \emph{\bibinfo{journal}{Phys. Rev. A}} \textbf{\bibinfo{volume}{74}},
  \bibinfo{pages}{042318} (\bibinfo{year}{2006}).

\bibitem{Schuster07}
\bibinfo{author}{Schuster, D.~I.} \emph{et~al.}
\newblock \bibinfo{title}{Resolving photon number states in a superconducting
  circuit}.
\newblock \emph{\bibinfo{journal}{Nature}} \textbf{\bibinfo{volume}{445}},
  \bibinfo{pages}{515--518} (\bibinfo{year}{2007}).

\bibitem{Roy05}
\bibinfo{author}{Roy, S.~M.}
\newblock \bibinfo{title}{Multipartite separability inequalities exponentially
  stronger than local reality inequalities}.
\newblock \emph{\bibinfo{journal}{Phys. Rev. Lett.}}
  \textbf{\bibinfo{volume}{94}}, \bibinfo{pages}{010402}
  (\bibinfo{year}{2005}).

\bibitem{Toth05}
\bibinfo{author}{T\'oth, G.} \& \bibinfo{author}{G\"uhne, O.}
\newblock \bibinfo{title}{Entanglement detection in the stabilizer formalism}.
\newblock \emph{\bibinfo{journal}{Phys. Rev. A}} \textbf{\bibinfo{volume}{72}},
  \bibinfo{pages}{022340} (\bibinfo{year}{2005}).

\end{thebibliography}


\begin{thebibliography}{1}
\expandafter\ifx\csname url\endcsname\relax
  \def\url#1{\texttt{#1}}\fi
\expandafter\ifx\csname urlprefix\endcsname\relax\def\urlprefix{URL }\fi
\providecommand{\bibinfo}[2]{#2}
\providecommand{\eprint}[2][]{\url{#2}}

\bibitem{DiCarlo09}
\bibinfo{author}{Di{C}arlo, L.} \emph{et~al.}
\newblock \bibinfo{title}{Demonstration of two-qubit algorithms with a
  superconducting quantum processor}.
\newblock \emph{\bibinfo{journal}{Nature}} \textbf{\bibinfo{volume}{460}},
  \bibinfo{pages}{240--244} (\bibinfo{year}{2009}).

\bibitem{Santavicca08}
\bibinfo{author}{Santavicca, D.} \& \bibinfo{author}{Prober, D.}
\newblock \bibinfo{title}{Impedance-matched low-pass stripline filters}.
\newblock \emph{\bibinfo{journal}{Meas. Sci. Technol.}}
  \textbf{\bibinfo{volume}{19}}, \bibinfo{pages}{087001}
  (\bibinfo{year}{2008}).

\end{thebibliography}

\end{document}

\title{Supplementary Material for\\`Preparation and Measurement of Three-Qubit Entanglement in a Superconducting Circuit'}
\author{L.\ DiCarlo}
\affiliation{Departments of Physics and Applied Physics, Yale University, New Haven, CT 06511, USA}
\author{M.\ D.\ Reed}
\affiliation{Departments of Physics and Applied Physics, Yale University, New Haven, CT 06511, USA}
\author{L.\ Sun}
\affiliation{Departments of Physics and Applied Physics, Yale University, New Haven, CT 06511, USA}
\author{B.\ R. \ Johnson}
\affiliation{Departments of Physics and Applied Physics, Yale University, New Haven, CT 06511, USA}
\author{J.\ M. \ Chow}
\affiliation{Departments of Physics and Applied Physics, Yale University, New Haven, CT 06511, USA}
\author{J.\ M. \ Gambetta}
\affiliation{Department of Physics and Astronomy and Institute for Quantum Computing, University of Waterloo, Waterloo, Ontario N2L 3G1, Canada}
\author{L.\ Frunzio}
\affiliation{Departments of Physics and Applied Physics, Yale University, New Haven, CT 06511, USA}
\author{S.\ M. \ Girvin}
\affiliation{Departments of Physics and Applied Physics, Yale University, New Haven, CT 06511, USA}
\author{M.\ H. \ Devoret}
\affiliation{Departments of Physics and Applied Physics, Yale University, New Haven, CT 06511, USA}
\author{R.\ J. \ Schoelkopf}
\affiliation{Departments of Physics and Applied Physics, Yale University, New Haven, CT 06511, USA}
\date{\today}
\maketitle

\bibliography{References_GHZ}

\onecolumngrid\par

\begin{figure}[hb!]
%\psfrag{A}[c][c][1.0]{$\pi$}
%\psfrag{t}[c][c][1.0]{$\tau$}
%\psfrag{Q}[c][c][1.0]{$\delta \Vtw$}
%\psfrag{C}[c][c][1.0]{$\tau$}
%\psfrag{X}[c][c][1]{$\Vo$}
\centering
\includegraphics[width=97mm]{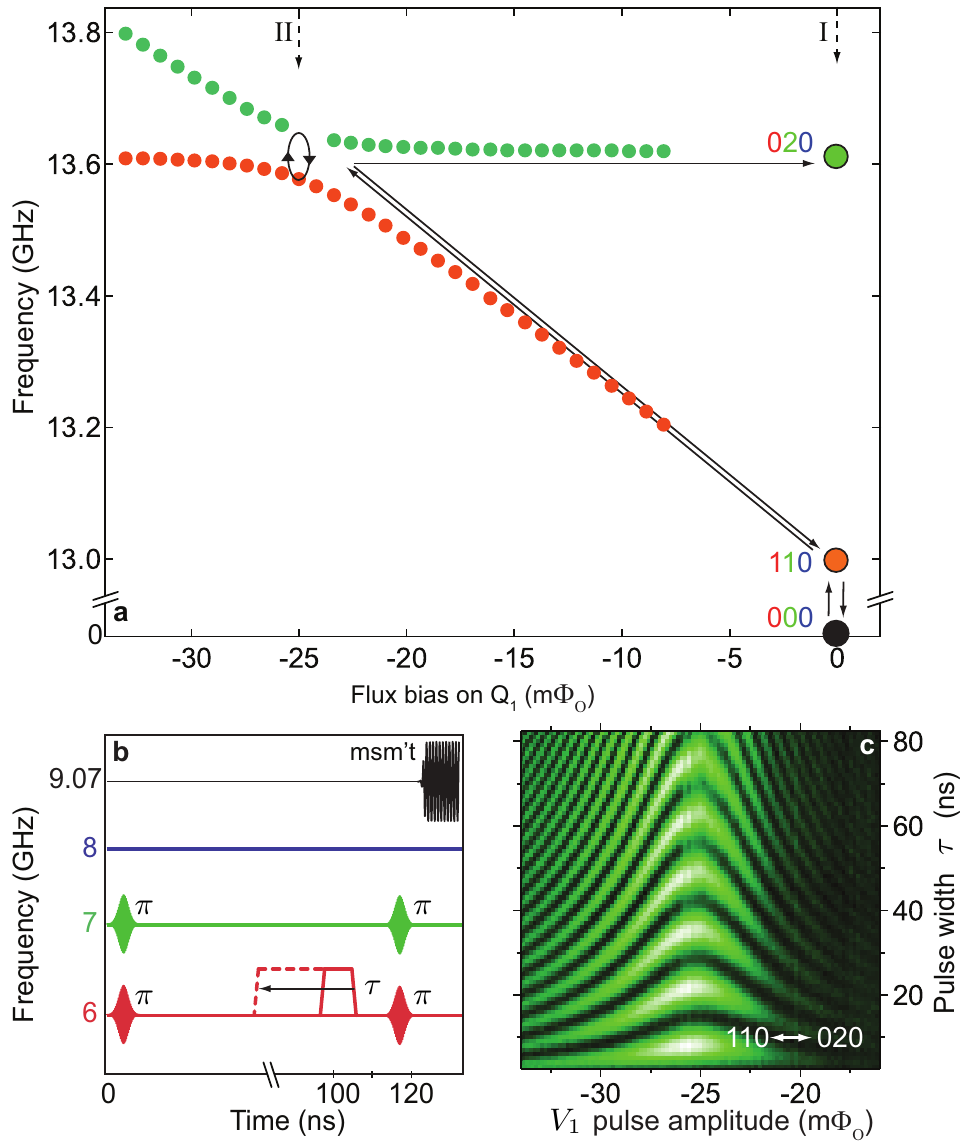}
\caption{{\bfseries Frequency- and time-domain characterization of the $\ket{110}\leftrightarrow\ket{020}$ avoided crossing.}
\textbf{a,}~Two-tone spectroscopy of computational level $\ket{110}$
and non-computational level $\ket{020}$ as a function of local bias on $\Qo$. The two levels become resonant at point II. Their cavity-mediated interaction produces a $73~\MHz$ splitting. This crossing is the primitive for $\Qo$-$\Qtw$ C-Phase gates.  \textbf{b,}~ Schematic of a pulse sequence, similar to that described in Fig.~2 of the main text, for calibrating the amplitude and duration $\tau$ of the required $\Vo$ pulse. \textbf{c,}~A full coherent oscillation on resonance takes $\tau=14~\ns$, consistent with the inverse of the minimum splitting in \textbf{a}, and setting the two-qubit gate time.
\label{fig:fig2}}
\end{figure}

\newpage
\vspace*{0pt plus 1fill}
\begin{figure}[h!]
\centering
\includegraphics[width=160mm]{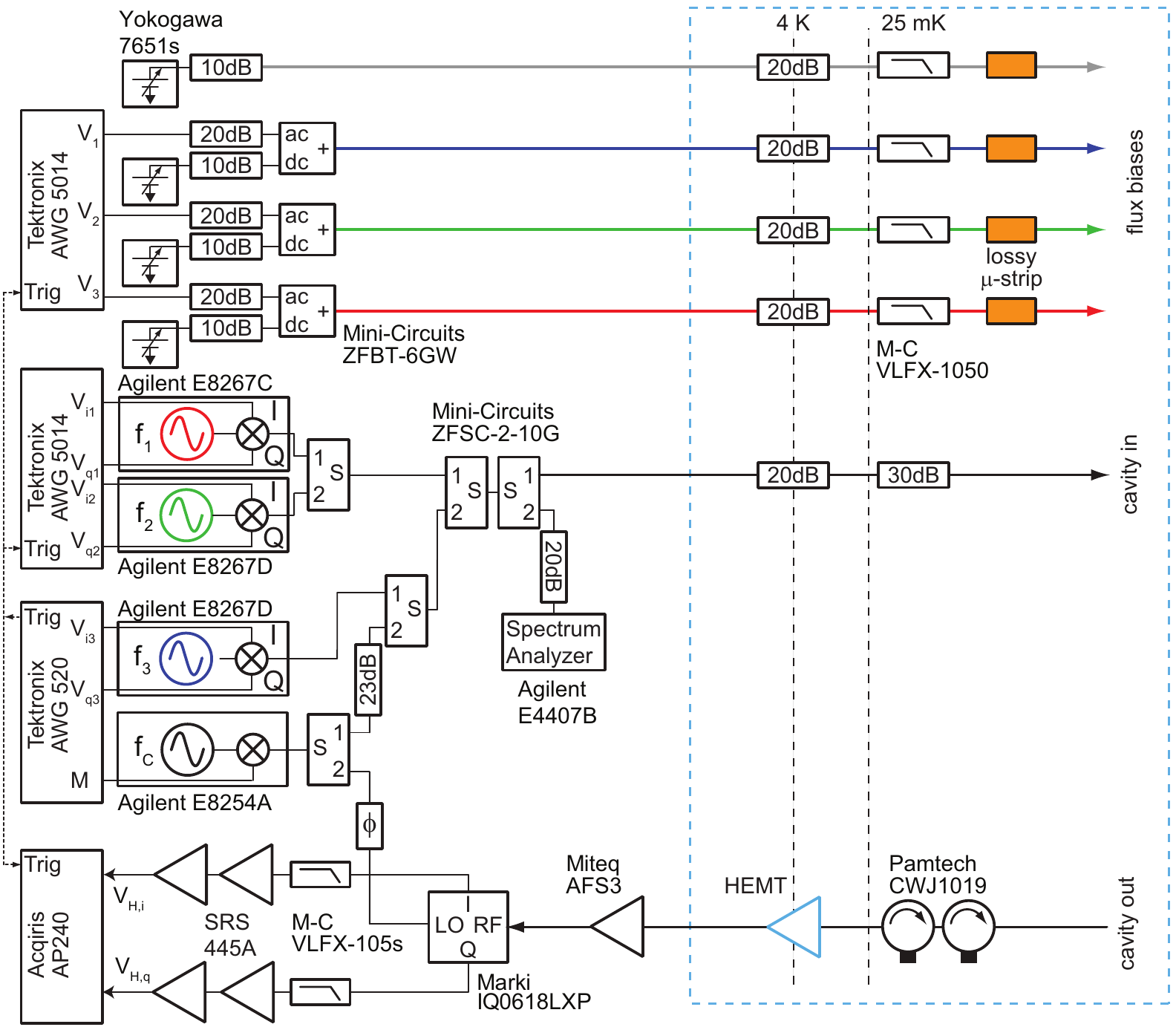}
\caption{{\bfseries Block diagram of room-temperature electronics and fridge wiring}.
Our setup is an expanded version of the one presented in Ref.~\onlinecite{DiCarlo09}, allowing full control of three qubits and biasing of a fourth. Microwave drives:  wideband I-Q modulated vector generators drive $x$- and $y$-rotations on qubits 1, 2, and 3, and arbitrary waveform generators (AWG, models 5014 and 520, running at 1~GSample/s with 14 and 10-bit vertical resolution) provide the 6 modulation envelopes.
A scalar generator is pulse modulated with an AWG marker to produce the measurement drive. Flux control: AWGs
produce voltage pulses $\Vo$ to $\Vth$ for fast flux biasing of $\Qo$ to $\Qth$. Yokogawa dc sources coupled via bias tees provide the quiescent bias (rather than the AWGs),  improving the frequency stability of the qubits when operated away from the flux sweet-spot (for example at point I) and permitting heavier filtering (attenuation) of the ac-coupled AWG flux drives. Inside the refrigerator, the combination of a reactive  low-pass filter and a lossy strip-line~\cite{Santavicca08} prevents spurious resonant qubit driving while keeping a nearly $50~\Ohm$ impedance at qubit transition frequencies (looking from the qubit side). Output amplification chain: The output line has  $\sim70~\dB$ gain and $\sim10~\K$ noise temperature  in the $4$--$8~\GHz$ range. An I-Q mixer and a two-channel averager ($2~\ns$, 8-bit sampling) are used for homodyne detection of the cavity quadratures.  We have eliminated the local oscillator used in Ref.~\onlinecite{DiCarlo09} by splitting the cavity drive. The AWGs, microwave synthesizers and acquisition card are clocked with a Rubidium frequency standard (SRS~FS725, not shown).
\label{fig:fig5}}
\end{figure}
\vspace*{0pt plus 1fill}

\begin{figure}[ht!]
%\psfrag{a}[l][b][0.8]{$\Rypt$}
%\psfrag{b}[c][b][0.8]{$R_{z}^{\theta_3}$}
%\psfrag{c}[c][b][0.8]{$R_{z}^{\theta_2}$}
%\psfrag{d}[c][b][0.8]{$R_{z}^{\theta_1}$}
%\psfrag{e}[c][b][0.8]{$12~\ns$}
%\psfrag{f}[c][b][0.8]{$14~\ns$}
%\psfrag{g}[c][b][0.8]{$5~\ns$}
%\psfrag{h}[l][b][0.8]{$\Rxpt$}
%\psfrag{j}[l][b][0.8]{$\Rxpt$}
%\psfrag{k}[l][b][0.8]{$\Rxp$}
\centering
\includegraphics[width=87mm]{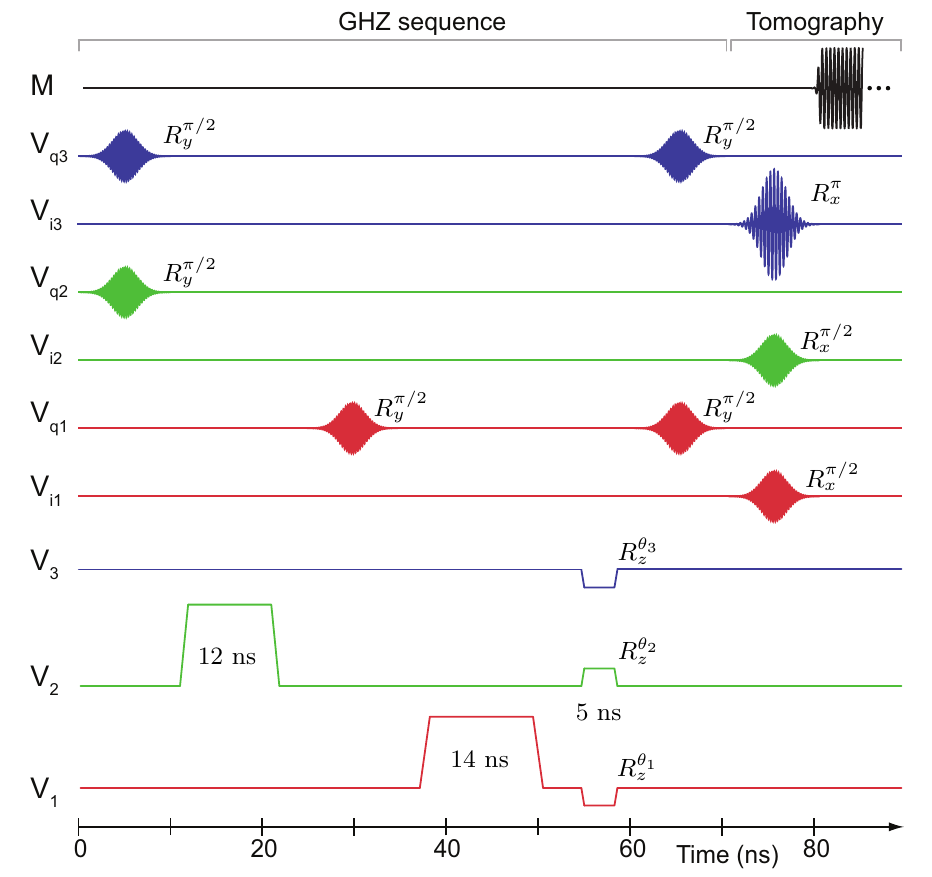}
\caption{
Illustration of the pulse sequence producing the GHZ state shown in Fig.~3c of the main text. The tomographic pre-rotations shown transform the measurement  into $\mVH=\beta_{ZII}\langle YII\rangle+\beta_{IZI}\langle IYI\rangle-\beta_{IIZ}\langle IIZ\rangle+\beta_{ZZI}\langle YYI\rangle-\beta_{ZIZ}\langle YIZ\rangle-\beta_{IZZ}\langle IYZ\rangle-\beta_{ZZZ}\langle YYZ\rangle$ (see Methods). All microwave pulses implementing the $x$- and $y$-rotations have Gaussian envelopes, with standard
deviation $\sigma=2~\ns$, truncated at $\pm 2 \sigma$. The rotation axis is set using I-Q (vector) modulation (see Fig.~\ref{fig:fig5}), and the rotation angle is controlled by pulse amplitude. To avoid overlap, a buffer of $3~\ns$ is inserted between all pulses. The full sequence (including tomography pre-rotations) is completed in $81~\ns$. The measurement tone is finally pulsed on for $1~\us$.
\label{fig:fig6}}
\end{figure}

\begin{figure*}[hb!]
%\psfrag{A}[c][b][1]{$\mathrm{Re}\left(\rho\right)$}
%\psfrag{B}[c][b][1]{$\mathrm{Im}\left(\rho\right)$}
\centering
\includegraphics[width=89mm]{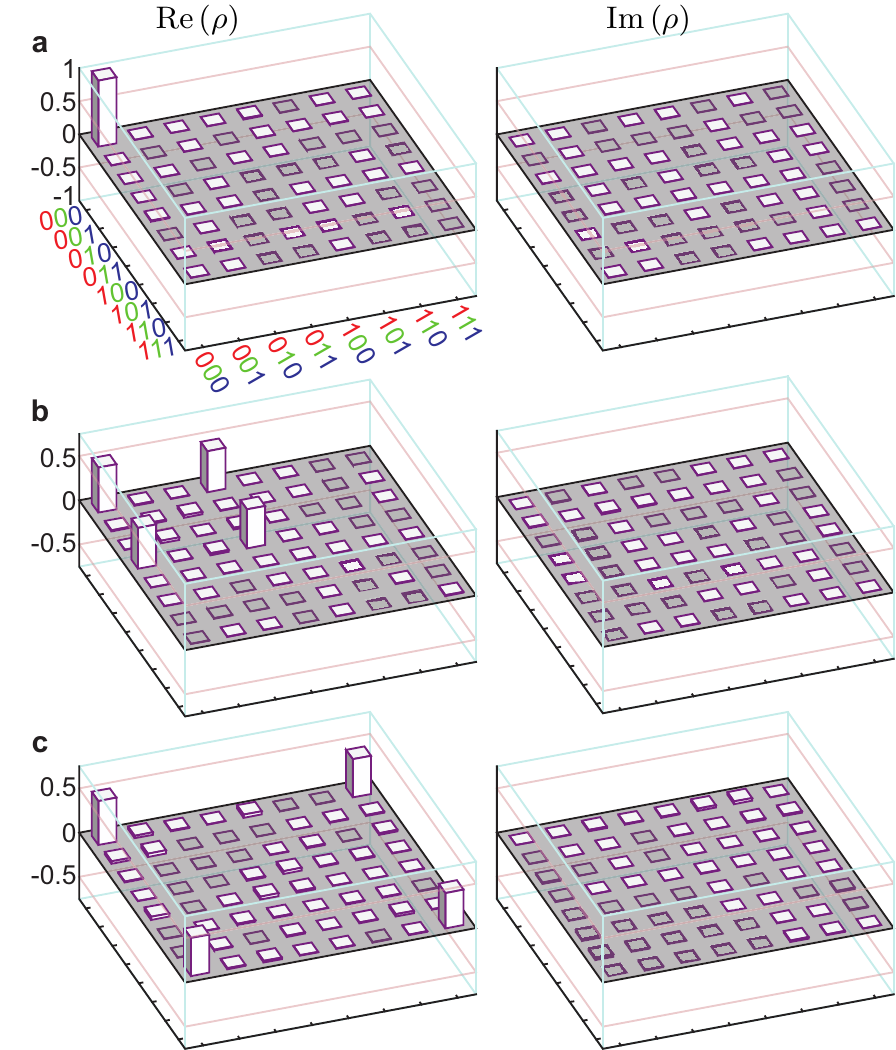}
\caption{{\bfseries Cityscape plots of reconstructed density matrices}.
The traditional visualization (also known as Manhattan plots) of the real and imaginary parts of the density matrices $\rho$ shown in Fig.~3 of the main text. Each $\rho$ is obtained from the corresponding Pauli set using Eq.~(1) in the main text. Note that maximum-likelihood estimation is not used to constrain $\rho$ to be positive semi-definite.
\label{fig:fig7}}
\end{figure*}